\documentclass[prl,twocolumn,showpacs,preprintnumbers,amsmath,amssymb]{revtex4}
\usepackage{graphicx}
\usepackage{dcolumn}
\usepackage{bm}

\begin{document}

\title{Two energy scales and close relationship between the pseudogap and superconductivity in underdoped cuprate superconductors}

\author{Hai-Hu Wen$^1$ and Xiao-Gang Wen$^2$}

\affiliation{$^1$National Laboratory for Superconductivity,
Institute of Physics and National Laboratory for Condensed Matter
Physics, Chinese Academy of Sciences, P. O. Box 603,Beijing 100080,
P. R..China}

\affiliation{$^2$Department of Physics, Massachusetts Institute of
Technology, Cambridge, Massachusetts 02139, USA}

\begin{abstract}
By measuring the low temperature specific heat, the low energy
quasi-particle excitation has been derived and analyzed in
systematically doped La$_{2-x}$Sr$_{x}$CuO$_{4}$ single crystals.
The Volovik's relation predicted for a $d$-wave superconductor has
been well demonstrated in wide doping regime, showing a robust
evidence for the $d$-wave pairing symmetry. Furthermore the nodal
gap slope $v_\Delta$ of the superconducting gap is derived and is
found to follow the same doping dependence of the pseudogap obtained
from ARPES and tunnelling measurement. This strongly suggests a
close relationship between the pseudogap and superconductivity.
Taking the entropy conservation into account, we argue that the
ground state of the pseudogap phase should have Fermi arcs with
finite density of states at zero K, and the transport data show that
it behaves like an insulator due to probably weak localization. A
nodal metal picture for the pseudogap phase cannot interpret the
data. Based on the Fermi arc picture for the pseudogap phase it is
found that the superconducting energy scale or $T_c$ in underdoped
regime is governed by both the maximum gap and the spectral weight
from the Fermi arcs. This suggests that there are two energy scales:
superconducting energy scale and the pseudogap. The
superconductivity may be formed by the condensation of Fermi arc
quasiparticles through pairing by exchanging virtue bosons.
\end{abstract}

\pacs{74.25.Bt, 74.25.Dw, 74.72.Dn}

\maketitle

Since the discovery of the cuprate superconductors, about 20 years
have elapsed without a consensus about its mechanism. Many exotic
features beyond the Bardeen-Cooper-Schrieffer theory have been
observed. One of the core issues is about the origin of the pseudoap
(PG)~\cite{Timusk99} and its relationship with the
superconductivity. One scenario assumes that the PG (with the energy
scale $\Delta_p$) marks only a competing or coexisting order with
the superconductivity and it has nothing to do with the pairing
origin. However other pictures, typically the Anderson's
resonating-valence-bond (RVB)~\cite{Anderson87} model predicts that
the spin-singlet pairing in the RVB state (which causes the
formation of the PG) may lend its pairing strength to the mobile
electrons and make them naturally pair and then condense at $T_c$.
According to this picture there should be a close relationship
between the PG and the superconductivity.

\begin{figure}
\includegraphics[width=9cm]{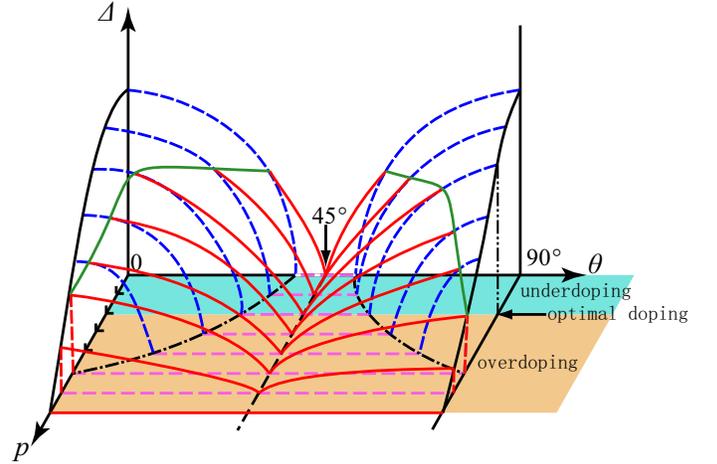}
\caption{Schematic plot for the general quasiparticle gap, the PG
energy (dashed blue line), superconducting energy scale (red solid
curve) and nodal gap slope at $\theta=45^\circ$. The angle $\theta$
counts from the $k$-space axis $k_x$, $p$ represents the hole
concentration.} \label{fig1}
\end{figure}

In order to get a deeper insight about the relationship between the
PG and superconductivity, we need to collect the information for the
PG and the superconducting energy scale, especially their doping
dependence. The PG values $\Delta_p$ or its corresponding
temperature $k_BT^\ast$ ($\propto\Delta_p$) and its doping
dependence have been measured through experiments. To determine the
superconducting energy scale, we note that the normal state Fermi
surface is formed by four small arcs near the nodal
points~\cite{Shen05}. As temperature is lowered below $T_c$, a new
gap opens on these arcs. To illustrate this point more clearly, in
Fig. 1 we present a schematic plot for different gaps or energy
scales. The dashed blue line represents the gap structure of the PG
state, assuming the presence of Fermi arcs near the nodal points.
The region of zero gap corresponds to the Fermi arc at zero K if the
superconductivity would be suppressed completely. The red solid line
represents the general quasiparticle (QP) gap on the Fermi arcs in
the superconducting state. In superconducting state, the general QPs
gap may construct a standard $d$-wave gap. Based on this picture, we
see that the nodal gap slope, which is defined as
$v_\Delta=\left|{d\Delta_s/d\theta}\right|_{\mathrm{node}}/\hbar
k_F$, can be used to determine the superconducting energy scale. The
relationship between $v_\Delta$ and the maximum PG $\Delta_p$
remains to be a big puzzle. In particular, the two quantities may be
independent of each other if the superconductivity is not induced by
the formation of the PG. Therefore to measure the nodal gap slope
$v_\Delta$ near nodal point in the zero temperature limit becomes
highly desired. When combined with the known results on the PG
$\Delta_p$, this will allow us to tell whether there is a
relationship between the PG and the superconductivity. In this
paper, we report the evidence of a proportionality between
$v_\Delta$ and the PG temperature $T^\ast$. We also find that $T_c$
is governed by both $v_\Delta$ and the spectral weight from the
Fermi arcs ($k_{\mathrm{arc}}$).

We determine the properties of the nodal quasiparticles by measuring
low temperature electronic specific heat on systematically doped
La$_{2-x}$Sr$_x$CuO$_4$ single crystals ($p=0.063$, 0.069, 0.075,
0.09, 0.11, 0.15, 0.22). Details about the sample characterization,
the specific heat measurement, the residual linear term and
extensive analysis were reported in our recent
papers~\cite{Wen04,Wen05}. Here we present a further analysis of
these data. The full squares in Fig. 4 represent the transition
temperatures of our samples. In all measurements the magnetic field
was applied parallel to $c$-axis.

\begin{figure}
\includegraphics[width=9cm]{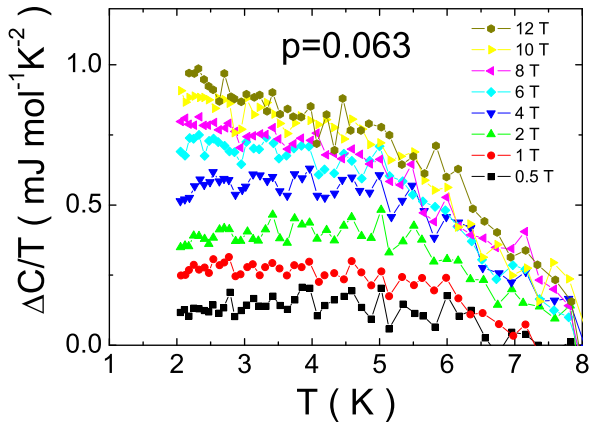}
\caption {The field induced change of the specific heat in low
temperature region for the very underdoped sample ($p=0.069$). One
can see that the field induces extra quasiparticle density of
states.} \label{fig2}
\end{figure}

One of the important discovery in our measurement is that the low
temperature specific heat coefficient $\gamma=C_e/T$ always
increases with the magnetic field. An example for the very
underdoped sample ($p=0.069$) is shown in Fig. 2. This behavior is
in sharp contrast with the low temperature thermal conductivity
data~\cite{Sutherland05,Sun04} which shows an unchanged or even
decline of the thermal conductivity $\kappa _0/T$. Our data clearly
show that the field has induced new quasiparticles, although they
are localized leading to the decrease of the thermal conductivity
and the diverging of the resistivity. In this sense the ground state
when the superconductivity is suppressed completely has finite
density of states at the Fermi level although it is insulating with
localized quasiparticles.

Next we show the robust evidence of $d$-wave pairng symmetry in
regimes from very underdoped to very overdoped. It has been widely
perceived that the pairing symmetry in the hole doped cuprate
superconductors is of $d$-wave with line nodes in the gap function.
In the mixed state, due to the presence of vortices,
Volovik~\cite{Volovik93} pointed out that supercurrents around a
vortex core lead to a Doppler shift to the QP excitation spectrum.
This will dominate the low energy QP excitation and the specific
heat (per mol) behaves as $C_{\mathrm{vol}}=AH^{1/2}$ with
$A\propto1/v_\Delta$. This square-root relation has been verified by
many measurements which were taken as evidence for $d$-wave
symmetry. In this way one can determine the nodal gap slope
$v_\Delta$. Since the phonon part of the specific heat is
independent on magnetic field, this allows to remove the phonon
contribution by subtracting the $C/T$ at a certain field with that
at zero field, one has $\Delta\gamma=\Delta C/T=[C(H)-C(0)]/T$. For
a $d$-wave superconductor, in the zero temperature limit
$\Delta\gamma=\Delta C/T=C_{\mathrm{vol}}/T=AH^{1/2}$ is
anticipated.

\begin{figure}
\includegraphics{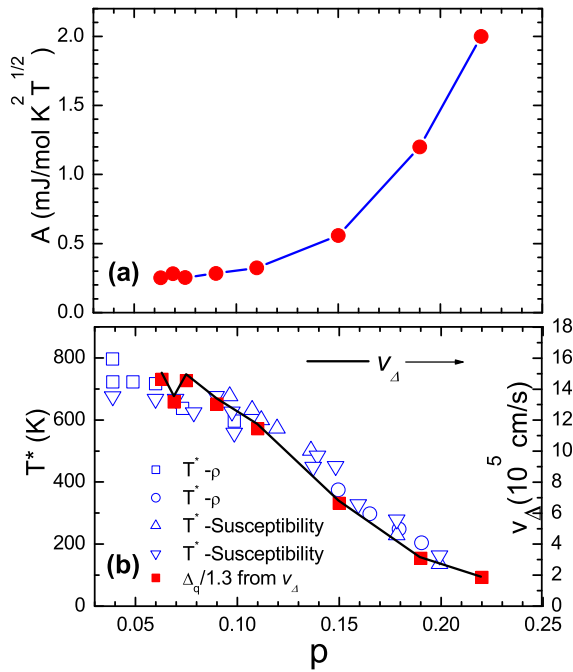}
\caption{(a) Doping dependence of the pre-factor $A$ determined in
present work (circles). Here the point at $p=0.19$ was adopted from
the work by Nohara \textit{et al.} on a single
crystal~\cite{Nohara00}. (b) Doping dependence of the PG temperature
$T^\ast$ (open symbols) summarized in Ref. 1 (see Fig. 26 there) and
our data $v_\Delta$ (solid line). The full squares represent the
calculated virtual maximum quasi-particle gap $\Delta_q$ derived
from $v_\Delta$ without any adjusting parameters. Surprisingly both
sets of data are correlated through a simple relation
$\Delta_q\approx0.46k_BT^\ast$ although they are determined in
totally different experiments. This result implies a close
relationship between the PG $\Delta_p$ and the nodal gap slope
$v_\Delta$.} \label{fig3}
\end{figure}

In order to get $\Delta\gamma$ in the zero temperature limit, we
extrapolate the low temperature data of $C/T$ vs. $T^2$ (between 2 K
to 4 K) to zero K. The data taken in this way and normalized at 12 T
are presented in our recent papers~\cite{Wen04,Wen05}. It is found
that the Volovik's $H^{1/2}$ relation describes the data rather well
for all doping concentrations. Furthermore we can determine the
prefactor $A$ in $\Delta\gamma=\Delta
C/T=C_{\mathrm{vol}}/T=AH^{1/2}$ and $v_{\Delta}$. Fig. 3(a) shows
the doping dependence of the pre-factor $A$. The error bar is
obtained by fitting the extracted zero temperature data to
$\Delta\gamma=AH^{1/2}$. For a typical $d$-wave superconductor, by
calculating the Dirac fermion excitation spectrum near the nodes, it
was shown that~\cite{Hussey02}
\begin{equation}
A=\alpha_p\frac{4k_B^2}{3\hbar l_c}\sqrt{\frac{\pi}{\Phi_0}}
\frac{nV_{\mathrm{mol}}}{v_\Delta} \label{eq1}
\end{equation}
here $l_c=13.28$ {\AA} is the $c$-axis lattice constant,
$V_{\mathrm{mol}}=58$ cm$^{3}$ (the volume per mol), $\alpha_p$ a
dimensionless constant taking 0.5 (0.465) for a square (triangle)
vortex lattice, $n=2$ (the number of Cu-O plane in one unit cell),
$\Phi _{0}$ the flux quanta. The $v_\Delta$ has then been calculated
without any adjusting parameter (taking $\alpha_p=0.465$) and shown
in Fig. 3(b). It is remarkable that $v_\Delta$ has a very similar
doping dependence as the PG temperature $T^\ast$, indicating that
$v_\Delta\propto T^\ast\propto\Delta_p$. If converting the data
$v_\Delta$ into the virtual maximum quasiparticle gap ($\Delta_q$)
via $v_\Delta=2\Delta_q/\hbar k_F$, here $k_F=\pi/\sqrt2a$ is the
Fermi vector of the nodal point with $a=3.8$ {\AA} (the in-plane
lattice constant), surprisingly the resultant $\Delta_q$ value
[shown by the filled squares in Fig. 3(b)] is related to $T^\ast$ in
a simple way ($\Delta_q\approx0.46k_BT^\ast$). It is important to
emphasize that this result is obtained without any adjusting
parameters. Counting the uncertainties in determining $T^\ast$ and
the value of $\alpha_p$, this relation is remarkable since
$\Delta_q$ and $T^\ast$ are determined in totally different
experiments. Because $v_\Delta$ (or $\Delta_q$) reflect mainly the
information near nodes which is predominantly contributed by the
superconductivity, above discovery, i.e., $v_\Delta\propto
T^\ast\propto\Delta_p$ (or $\Delta_q=0.46k_BT^\ast$) strongly
suggests a close relationship between the PG and superconductivity.

In above discussion, we see the consistency between our low
temperature specific heat data and the Volovik's relation
$\Delta\gamma=A\sqrt{H}$. This seems surprising since the
temperature range considered here is about several Kelvin. At such
an energy scale, the impurity scattering will strongly alter the DOS
in the low energy region. However, by applying a magnetic field, the
Doppler shift of the quasiparticle excitation spectrum will
contribute a new part to DOS. As argued in our recent paper, this
energy shift can be described by $\Delta\varepsilon=3.67\alpha
_{FL}\sqrt{B/1T}$ meV. For example, taking the maximum field (12 T)
in our experiment, we get $\Delta\varepsilon=12.2\alpha _{FL}$ meV
which is actually a relatively large energy scale compared to the
temperature since $\alpha_{FL}\approx1$~\cite{Wen05}. This may
explain why the Volovik's simple relation $\Delta\gamma=A\sqrt{H}$
can be easily observed in our single crystals with inevitable
certain amount of impurities.

\begin{figure}
\includegraphics[width=9cm]{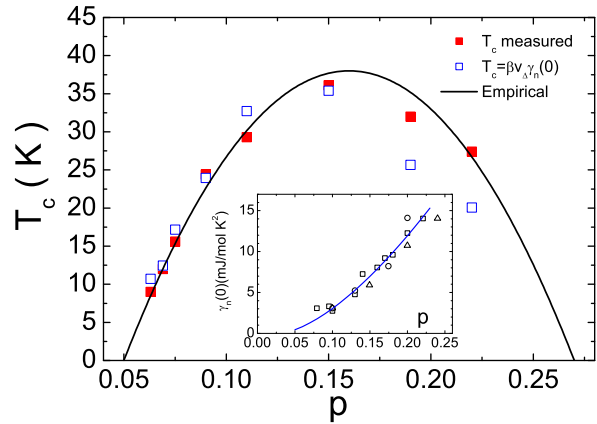}
\caption {Doping dependence of the measured $T_c$ (full squares) and
that calculated one by $T_c=\beta v_\Delta\gamma_n(0)$ (open
squares) with $\beta=0.7445$ K$^{3}$ mol~s/J~m. The solid line
represents the empirical relation
$T_c=T_c^{\mathrm{max}}=1-82.6(p-0.16)^2$ with
$T_c^{\mathrm{max}}=38$ K. The inset shows the residual value of
$\gamma_{n}(0)$ of the PG state, the solid line is a fit to
$\gamma_n=\zeta(p-p_c)^\eta$ with $\zeta=182.6$, $p_c=0.03$, and
$\eta=1.54$.} \label{fig4}
\end{figure}

In the following we will investigate what governs $T_c$. Bearing the
doping dependence of $v_\Delta$ in mind, it is easy to understand
that $v_\Delta\hbar k_F$ should not be a good estimate of the
superconducting energy scale for the underdoped samples since the
$T_c$ and $v_\Delta$ have opposite doping dependence. The basic
reason is that the normal-state Fermi surface contains small arcs of
length $k_{\mathrm{arc}}$ near the nodal points. The superconducting
transition occurs by forming extra gaps on these Fermi arcs. So the
effective superconducting energy scale should be estimated as
$E_s\approx1/2v_\Delta\hbar k_{\mathrm{arc}}$. From the normal state
electronic specific heat $C_{el}=\gamma_nT$, we have
$\gamma_n=4nk_B^2k_{\mathrm{arc}}V_{\mathrm{mol}}/\hbar v_Fl_c$.
Assuming $E_s\approx k_BT_c$ we find
\begin{equation}
T_c=\alpha_s\frac{\hbar
v_Fl_c\gamma_nv_\Delta}{8nk_B^3V_{\mathrm{mol}}}=\beta\gamma_nv_\Delta
\label{eq2}
\end{equation}
where $\alpha_{s}$ is a dimensionless constant in the order of
unity, $v_F$ is the nodal Fermi velocity normal the Fermi surface.
The value of $\gamma _n(0)$ can be estimated from specific
heat~\cite{Matsuzaki04}. Here we take the values for $\gamma_n(0)$
summarized by Matsuzaki \textit{et al.}~\cite{Matsuzaki04} and fit
it (in unit of mJ/mol K$^2$) with a formula
$\gamma_n=\zeta(p-p_c)^\eta$ yielding $\zeta=182.6$, $p_c=0.03$,
$\eta=1.54$. In the inset of Fig. 4 we present the doping dependence
of the zero-temperature specific heat coefficient $\gamma_n(0)$ from
which one can calculate $k_{\mathrm{arc}}$. In the main frame of
Fig. 4 we present the doping dependence of the measured $T_{c}$
(filled squares) and the calculated value (open squares) by eq. (2)
with $\beta=0.7445$ K$^3$ mol~s/J~m. In underdoped region, the
measured and calculated $T_c$ values coincide rather well implying
the validity of eq. (2). So the energy scale of the
superconductivity is not given by $v_\Delta\hbar k_F$, but by
$E_s=1/2v_\Delta\hbar k_{\mathrm{arc}}$ or more precisely by eq. (2)
in the underdoped region. In overdoped region, $\gamma _n(0)$ will
gradually become doping independent, therefore one expects
$T_c\propto\Delta_q\propto v_\Delta$.

\begin{figure}
\includegraphics[width=12cm]{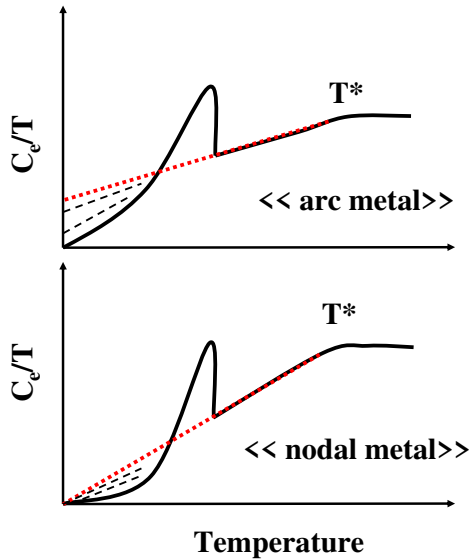}
\caption {Schematic plot for temperature dependence of the
electronic specific heat coefficient $C_e/T$ for (a) ``arc metal"
with a finite DOS at $T=0$ and (b) ``nodal metal" for the PG phase.
By applying a magnetic field, in the superconducting state at $T=0$,
it is found that the DOS increases, which is not consistent with the
expectation of the nodal metal ground state for the PG phase. The
two short dashed lines in the zero temperature limit illustrate how
does $C_e/T$ change with increasing magnetic field (from bottom to
up) if the PG ground state is (a) an ``arc metal" or (b) a ``nodal
metal".} \label{fig5}
\end{figure}

Our discussion here is totally based on the assumption of the Fermi
arc ground state for the PG phase. Although some preliminary
evidence for the existence of Fermi arcs has been found by
ARPES~\cite{Shen05}, in the following we argue that the Fermi ``arc
metal" instead of the ``nodal metal" is the ground state of the PG
phase. From the specific heat data of Matsuzaki \textit{et
al.}~\cite{Matsuzaki04} and Loram \textit{et al.}~\cite{Loram01} one
can see that the PG phase (when the superconductivity is completely
suppressed) has actually a finite DOS at zero K based on the entropy
conservation consideration, as shown by Fig. 5(a). This finite DOS
at zero K of the PG phase can be interpreted as due to two possible
reasons: either induced by the impurity scattering of the nodal
quasiparticles of a $d$-wave PG (if it would be a ``nodal metal"),
or given by the zero-temperature Fermi arcs of the PG. For a ``nodal
metal" ground state, no extra quasiparticles can be generated by
increasing the magnetic field in the zero temperature limit (as
shown in Fig.5 (b)). However as shown in Fig. 2, our specific heat
data clearly show that there are extra DOS generated by applying the
magnetic field which is just the case shown by Fig. 5(a). Therefore
from experiments, it is evident that the specific heat behaves in
the way as Fig. 5(a) for an ``arc metal" instead of Fig. 5(b) for a
``nodal metal". It is thus tempting to conclude that there are Fermi
arcs near nodes in the PG phase in the zero temperature limit.

In summary, a close relationship between the PG and
superconductivity has been found. Based on the Fermi arc picture of
the PG phase, it is found that the superconducting energy scale in
underdoped regime is governed by both the maximum gap and the
spectral weight from the Fermi arcs. This suggests that there are
two energy scales and the superconductivity may be formed by the
Fermi arc quasiparticles through pairing mediated by exchanging
virtue bosons, such as the spin interaction or phonons.

\begin{acknowledgments}
This work is supported by the NSFC, the MOST 973 project
(2006CB601002), the CAS project (ITSNEM). We are grateful for
samples from F. Zhou at IOP (CAS) and Y. Ando from CRIEP (Tokyo,
Japan).

\end{acknowledgments}

\end{document}